\documentclass[]{aa}

\usepackage{graphics}
\usepackage{times}

\begin{document}
\thesaurus{7   % Stellar Atmospheres
(02.01.4;      % atomic processes
 02.12.3;      % line: profiles
 08.01.3)}     % stars: atmospheres

\title{Self-broadening in Balmer line wing formation in stellar atmospheres} 

\author{P. S. Barklem\inst{1} \and N. Piskunov\inst{1} \and B. J. O'Mara\inst{2}}
 
\offprints{P. S. Barklem (barklem@astro.uu.se)}

\institute{Uppsala Astronomical Observatory, Box 515, S 751-20 Uppsala, Sweden 
\and
	   Department of Physics, The University of Queensland, St Lucia, 4072, Australia}

\date{Received 7 July 2000 / Accepted 2 October 2000}

\maketitle

\begin{abstract}

Details of a theory of self-broadening of hydrogen lines are presented.  The main features of the new theory are that the dispersive-inductive components of the interaction (van der Waals forces) have been included, and the resonance components have been computed by perturbation theory without the use of the multipole expansion.  The theory is applied to lower Balmer lines and the theoretical and observational impact of the new broadening theory is examined.  It is shown that this theory leads to considerable differences in the predicted line profiles in cool stars when compared with previous theories which include only resonance interactions. In particular, the effect is found to be very important in metal poor stars. The theory provides a natural explanation for the behaviour of effective temperatures derived from Balmer lines by others using a theory which includes only resonance broadening. When applied to Balmer lines in the solar spectrum the theory predicts an improved agreement between observed and computed profiles for models which also match limb darkening curves and rules out a model which does not. However significant discrepancies still remain which could be due to inadequacies in our theory or the atmospheric model or both.

\end{abstract}

\section{Introduction}

Hydrogen line wings are one of the strongest tests of model stellar atmosphere structure. In the majority of stars hydrogen is a dominant source of continuous opacity and thus for strong hydrogen lines the abundance parameter is excluded.  In hot stars, where the broadening of hydrogen lines is dominated by protons and electrons produced by the ionisation of hydrogen, the hydrogen line profiles are dependent only on atmospheric structure and properties of the hydrogen atom.  In cool stars the metallicity is important as ionisation of metals is the principal source of ions and electrons which contribute to the Stark broadening of the lines while hydrogen atoms in their ground state produce self-broadening of the lines.  Further, hydrogen lines can be observed in all stars, unlike for example limb darkening curves which are one of the strongest tests for solar photosphere models.  The large range of opacities within a single line means that many different depths are probed.

If the behaviour of the hydrogen atom in the conditions of stellar atmospheres is understood, the hydrogen absorption lines can be a powerful diagnostic.  The transition probabilities are known with extremely high accuracy.  Furthermore, the line profile shape is particularly sensitive to atmospheric structure, due to the unique situation of the broadening which derives from the ``accidental degeneracy'' of states in the hydrogen atom.  However, this degeneracy means that the broadening is extremely complex by comparison with metallic lines.

Recently there have been a number of applications of hydrogen lines, particularly lower Balmer lines, to the analysis of stellar photospheric models, and in particular models of stellar convection (Fuhrmann~et~al.~\cite{fuhrmann1,fuhrmann2}, Van't Veer-Menneret \& M\'egessier~\cite{vanV}, Castelli~et~al.~\cite{castelli}, Gardiner~et~al.~\cite{gardiner}).  Of course, such analyses will be dependent on the accuracy of the theories describing the hydrogen atom properties.  Of particular importance for analyses of photospheres is the broadening of the wings in stellar photospheric conditions.  This is especially true when testing convection theories, as convection affects the atmosphere in the deeper layers which do not contribute to the core of the line. 

It was pointed out by Lortet \& Roueff~(\cite{lortet_roueff}) that the effect of neglecting dispersive-inductive forces should be significant, however, this seems to have gone largely unnoticed. They showed that relative to resonance broadening, dispersive interactions make a significant contribution to the self-broadening of Balmer lines.  For Paschen lines they demonstrated that dispersive interactions should dominate the self-broadening.  However, in their analysis they used an inadequate theory of the dispersive-inductive interaction which is known to underestimate this type of broadening by typically a factor of two. In this paper a theory of self-broadening of hydrogen lines which includes a better treatment of both resonance and dispersive-inductive interactions, which was announced in an earlier letter (Barklem~et~al.~\cite{bpo:let}, hereafter Paper I), is presented. It is shown that the inclusion of these interactions has a significant effect on the predicted profiles of Balmer lines in cool stars.

\section{Hydrogen Line Wing Broadening Mechanisms}

Hydrogen lines are broadened by a number of different mechanisms. In the wings of the lines the collisional processes dominate the line profile shape.   The more important of the known broadening mechanisms for hydrogen line wings in stellar atmospheres are
\begin{itemize}
\item quasistatic broadening by collisions with ions/protons
\item impact broadening by collisions with electrons
\item impact broadening by collisions with hydrogen atoms
\item radiative broadening
\item impact broadening by collisions with helium atoms
\item impact broadening by collisions with hydrogen molecules
\end{itemize}
The first three of these are expected to dominate with relative contributions depending on the effective temperature and metallicity of the star, and the particular line under study.

The quasistatic ion field splits the line up into Stark components which are broadened by collisions with fast moving electrons, hydrogen atoms and helium atoms.  As they carry an overall electric charge, electrons are substantially more effective at broadening than hydrogen atoms. However, in cool stars like the sun, with a solar composition, hydrogen atoms outnumber electrons by typically four orders of magnitude and by perhaps six orders of magnitude if the star is metal deficient. This results in collisions with hydrogen atoms being very important. In cool stars, helium atoms usually have a number density about an order of magnitude less than hydrogen atoms, they have no resonance interaction with hydrogen and have only half the polarisability and speed of hydrogen atoms so their contribution to the broadening is relatively unimportant. Broadening by collisions with hydrogen molecules will only become important
in very cool stars which are not considered here.

\section{Unified Stark Broadening Theories}

Due to it's importance in later analysis and discussions, we first make a brief summary of the situation for the Stark broadening, which includes the effects of the ion field and the electrons.  Stark broadening of hydrogen lines has received considerable attention.  The most used of the treatments are those of Vidal~et~al.~(\cite{vcs:theory,vcs:codes,vcs:tables}), and more recently Stehl\'e~(\cite{stehle:94}).
The theory of Vidal~et~al.~(\cite{vcs:theory}) was the first ``unified theory'' of electron and ion collisional broadening of hydrogen lines.  This theory was for the first time capable of computing the line profile over the whole line, from the impact limit at line centre to the quasistatic limit in the line wings, including the transition region.
The calculations of Stehl\'e~(\cite{stehle:94}) are based on the theory of Frisch \& Brissaud~(\cite{fb:mmm1,fb:mmm2}), the so called Model Microfield Method (MMM).  The major advantage of these calculations, over the Vidal~et~al.~(\cite{vcs:tables}) calculations, is the modelling of the dynamics of ions and non-Markovian effects (overlapping collisions).  As stated by Stehl\'e \& Jacquemot~(\cite{stehle_jacquemot}), ``MMM can be seen as an interpolating formalism between well described asymptotic 'static' and 'impact' limits''.  By comparison the Vidal~et~al.~(\cite{vcs:theory}) theory is simply a unified treatment of the two limits.

The Vidal~et~al.~(\cite{vcs:tables}) and Stehl\'e~(\cite{stehle:94}) calculations are in reasonable agreement, mostly showing differences in the line core due to ion dynamics.  In cool stars the ions are produced by the ionisation of heavy metals like iron which move relatively slowly so the effects of ion motion will be very small.  The Stehl\'e calculations show quite good agreement with experiment~(Stehl\'e~\cite{stehle:94}), noticeably better than Vidal~et~al.~(\cite{vcs:tables}) in the line core.  However, this difference is often lost in astrophysical applications when the profiles are folded with Doppler profiles (Lemke~\cite{lemke}).  Both theories show reasonable agreement in the wings.

\section{Self-Broadening Theory}

In the case of hydrogen, self-broadening refers to the broadening of hydrogen lines by collisions with other neutral hydrogen atoms.  It has been known for some time, that similar atoms undergo a resonance interaction when the states of the two atoms are capable of optical combination (Eisenschitz \& London~\cite{eisenlond}).  It is also well known that two neutral atoms have dispersive and inductive interactions, often called the van der Waals interaction.  The dispersive interaction corresponds to the simultaneous fluctuation of the atoms brought about by the repulsive electrostatic interaction of the electrons in each atom which promotes and demotes the electrons to virtual states. At long range the dispersive interaction dominates and a multipole expansion of the electrostatic interaction is valid whose first term leads to an interaction of the form $C_6/R^6$. There are also components of the interaction which correspond to the induction of virtual transitions in one atom due to the static field of the other, such interactions are usually less important.

We now outline a new theory of self-broadening in the impact approximation which includes both resonance and dispersive and inductive interactions, in a single theory.

\subsection{Overlapping Lines in the Impact Approximation}

As already explained, in stellar spectra the hydrogen lines are split into components by the quasistatic ion field and these components are then impact broadened by electrons, hydrogen atoms, and to a negligible extent, helium atoms. The combined impact broadening of the Stark components by electrons and hydrogen can be handled in a consistent way by the use of overlapping line theory.

Baranger~(\cite{baranger:overlap}) was first to examine the problem of pressure broadening of overlapping lines in the impact approximation.  The review by Peach~(\cite{peach}) covers many aspects of line broadening theory including the case of overlapping lines.  Adopting the notation of Peach the line shape for a transition between states of principle quantum number $n_i$ and $n_j$ is given by
\begin{eqnarray}
 \lefteqn{L(\omega) = \frac{1}{\pi} \mbox{Re} \sum_{\ell_i\ell_j\ell_i^\prime \ell_j^\prime} \langle\langle n_i \ell_i (n_j\ell_j)^*||\mbox{\boldmath$\mathrm{\delta}$}|| n_i \ell_i^\prime (n_j\ell_j^\prime)^* \rangle\rangle } \nonumber \\
&&\times \langle\langle n_i \ell_i^\prime  (n_j\ell_j^\prime)^* || \left[ \mbox{\boldmath$\mathrm{\mathit{h}}$} - i(\omega - \mbox{\boldmath$\mathrm{h_0}$}/\hbar)\right]^{-1} || n_i \ell_i (n_j\ell_j)^* \rangle\rangle
\label{eq:lineshape}
\end{eqnarray}
in the reduced line/doubled-atom space, with $\mbox{\boldmath$\mathrm{\delta}$}$ the operator corresponding to the electric dipole operator, and  $\mathbf{h_0}$ the Hamiltonian. The operator $\mbox{\boldmath${\mathit{h}}$}$ is the operator corresponding to $N\{1-\mathbf{S}_i\mathbf{S}_j^\dagger\}_{\mathrm{av}}$ in line space, where $N$ is the perturber density and $\{\}_{\mathrm{av}}$ indicates averaging over all possible orientations of the collision.  $\mathbf{S}$ is the collision or scattering matrix, and here the subscript refers to the two different upper and lower state subspaces.  

The dipole operator determines the strength of each component's contribution to the complete line, including possible interference between components.  In the reduced line space the matrix elements of $\mbox{\boldmath$\mathrm{\delta}$}$ are related to the reduced state space matrix elements by
\begin{eqnarray}
\lefteqn{\langle\langle n_i \ell_i (n_j\ell_j)^*||\mbox{\boldmath$\mathrm{\delta}$}|| n_i \ell_i^\prime (n_j\ell_j^\prime)^* \rangle\rangle } \hspace{30mm}\nonumber \\
&& = \langle n_i \ell_i ||\mathbf{d}||n_j \ell_j\rangle \langle n_i \ell_i^\prime ||\mathbf{d}||n_j \ell_j^\prime \rangle
\end{eqnarray}
where via the Wigner-Eckhart theorem (eg. Edmonds~\cite{edmonds}) one can find
\begin{eqnarray}
\langle n_i \ell_i ||\mathbf{d}||n_j \ell_j\rangle &=& (-1)^{\ell_i} \left[(2\ell_i+1)(2\ell_j+1)\right]^{\frac{1}{2}} \nonumber \\
&& \times \left( \begin{array}{ccc} \ell_i & 1 & \ell_j \\ 0 & 0 & 0 \end{array} \right)  \langle n_i \ell_i | r | n_j \ell_j \rangle
\end{eqnarray}
where $\langle n_i \ell_i | r | n_j \ell_j \rangle$ is now simply the radial component which can be computed by standard methods such as those discussed by Condon \& Shortley~(\cite{condon_shortley}).  We use the readily available computer code from Vidal~et~al.~(\cite{vcs:codes}).

\subsection{Semi-classical Treatment}

The second matrix in Eq.~(\ref{eq:lineshape}) determines the line profile shape characteristics for each component. 
In the semi-classical theory, assuming straight line trajectories we can show that
\begin{eqnarray}
\lefteqn{\langle\langle n_i \ell_i^\prime (n_j\ell_j^\prime)^* || \left[ \mbox{\boldmath$\mathrm{\mathit{h}}$} - i(\omega - \mathbf{h_0}/\hbar) \right] || n_i \ell_i (n_j\ell_j)^* \rangle\rangle = }  \nonumber \\
&&  N \int_0^\infty v f(v) \: \mathrm{d}v \int_0^\infty 2 \pi b \: \mathrm{d}b \nonumber \\
&& \times \left[ 1- \sum_{m_i m_j m_i^\prime m_j^\prime \mu}
\left( \begin{array}{ccc} \ell_j & 1 & \ell_i \\ -m_j & \mu & m_i \end{array} \right) \left( \begin{array}{ccc} \ell_j^\prime & 1 & \ell_i^\prime \\ -m_j^\prime & \mu & m_i^\prime \end{array} \right)   \right.\nonumber \\
&& \left. \langle n_i \ell_i^\prime m_i^\prime |\mathbf{S}_I| n_i \ell_i m_i \rangle \langle n_j \ell_j^\prime m_j^\prime |\mathbf{S}_J^\dagger| n_j \ell_j m_j \rangle \right] \nonumber \\
&&  - \delta_{\ell_i^\prime \ell_i} \delta_{m_i^\prime m_i} \delta_{\ell_j^\prime \ell_j} \delta_{m_j^\prime m_j} i (\omega - \omega_{ij}) 
\end{eqnarray}
where $b$ is the impact parameter of the collision, and $f(v)$ the distribution of velocities $v$.
This matrix is a complex square matrix of order $n_i n_j$, and once computed can be inverted easily by standard numerical techniques to give the matrix required in Eq.~(\ref{eq:lineshape}).

Determination of the $\mathbf{S}$ matrices can be simplified if it is assumed that there are no $\ell$-changing collision-induced transitions so that the $\mathbf{S}$ matrices are block diagonal.  We compute these matrices via the method proposed by Roueff~(\cite{roueff:74}) which accounts for changes in the orientation of the atoms during the collision relative to the single orientation in which the potentials are computed, via the $\mathbf{S}$ matrix.  The relevant expressions for the evolution have been presented in Anstee \& O'Mara~(\cite{ao:nad}) and Barklem \& O'Mara~(\cite{bo:pd}).

In our treatment of the broadening of metallic lines it is assumed that there are no collision-induced transitions, an assumption which is justified by the collisions being too slow. In hydrogen, due to the
accidental near $\ell$ degeneracy, this assumption may break down. This is discussed in Sect.~\ref{sect:valid}.

\subsection{The Interaction Potential}

\begin{figure}
%\vspace{5mm}
\setlength{\unitlength}{0.6cm}
\centerline{
\begin{picture}(7,6)(0,-0.5)
 \put(7,1){\circle*{0.5}}
 \put(1,1){\circle*{0.5}}
 \put(1,1){\line(1,0){6}}
 \put(1,1){\line(1,2){1}}
 \put(7,1){\line(-1,2){2}}
 \put(1,1){\line(1,1){4}}
 \put(7,1){\line(-5,2){5}}
 \put(2,3){\line(3,2){3}}
 \put(6.2,3){$p_2$}
 \put(1.0,2.1){$r_1$}
 \put(4,0.4){$R$}
 \put(3.2,3){$r_2$}
 \put(3.0,4.2){$r_{12}$}
 \put(4.2,2.3){$p_1$}
 \put(-2.5,0.0){perturber proton}
 \put(-2.1,3.0){perturber electron}
 \put(5.0,0.0){perturbed atom proton}
 \put(5.2,5){optical electron}
 \put(5,5){\circle*{0.1}}
 \put(2,3){\circle*{0.1}}
\end{picture}}
\caption{Model for electrostatic interaction between the two hydrogen atoms.}
\label{fig:atoms}
\end{figure}
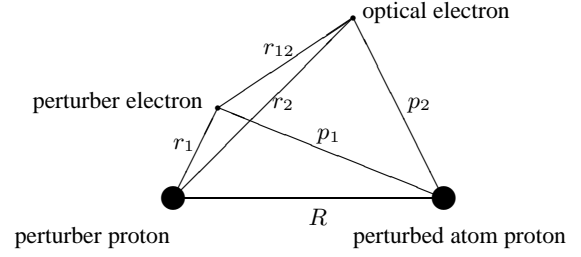

We model the interaction between two hydrogen atoms as shown in Fig.~\ref{fig:atoms}.  With reference to Fig.~\ref{fig:atoms} the electrostatic interaction is
\begin{equation}
V = \frac{1}{R}+\frac{1}{r_{12}}-\frac{1}{r_2}-\frac{1}{p_1} .
\label{eq:V}
\end{equation}

\begin{table}
\begin{center}
\begin{tabular}{cc}
\hline
State        &  $C_3$ (a.u.)      \\
\hline
1s2p$\sigma$ & $-1.1098579$   \\
1s2p$\pi$    & $0.5549290$    \\
1s3p$\sigma$ & $-0.1779785$   \\
1s3p$\pi$    & $0.0889893$    \\
1s4p$\sigma$ & $-0.0618475$   \\
1s4p$\pi$    & $0.0309238$    \\
1s5p$\sigma$ & $-0.0290382$   \\
1s5p$\pi$    & $0.0145191$    \\
\hline
\end{tabular}
\end{center}
\caption{The computed values of $C_3$ for resonance interactions due to ground state hydrogen perturbers.}
\label{tab:c3s}
\end{table}

The resonance interaction occurs between like atoms due to a possible exchange of excitation.  When one considers the interaction of a ground state hydrogen atom with an excited hydrogen atom the states $|100\rangle|n\ell m\rangle$ and $|n\ell m\rangle|100\rangle$ are degenerate.  If the excited state is a p state it has an allowed dipole transition to the ground state and the off-diagonal elements are quite large compared to the diagonal matrix elements of $V$ which at large separations can be neglected. It can be shown (see for example Margenau \& Kestner~\cite{margenau_kestner}, Fontana~\cite{fontana}) that the strength of the resonance interaction $\Delta E_{n1m,100}$ to first order is given by the matrix elements of $V$ between these two states. 
Analytic expressions for these matrix elements, to within a final numerical integration over the radial co-ordinate have been calculated from $V$ in Eq.~(\ref{eq:V}) with the assistance of the Mathematica package without resort to a multipole expansion of $V$. These matrix elements must, at long range, reduce to the form
\begin{equation}
\Delta E_{n1m,100} \approx \frac{C_3(n,m)}{R^3}
\end{equation}
where $C_3$ is a constant that must be computed.  Table~\ref{tab:c3s} shows values computed for $C_3$ for lower lying states of hydrogen interacting with a ground state perturber.  These were computed by direct evaluation of the dipole-dipole matrix element integrals using Mathematica.  The $n=2$ and $n=3$ elements were computed entirely analytically.  The higher states required some numerical evaluation.  These values are in excellent agreement with Stephens \& Dalgarno~(\cite{stephens}) and Kolos~(\cite{kolos}) for the 1s2p and 1s3p interactions and with our calculations based on an unexpanded $V$ when the interatomic separation is large. Note that the interaction can be of either sign and declines dramatically with increasing principal quantum number which results in a multipole expansion of $V$ being invalid for all but 2p states when employed in line broadening calculations.

For those perturbed atom states which are not connected to the ground state by an allowed dipole transition, the off diagonal matrix elements of $V$ (corresponding to forbidden transitions) are zero or negligible. Thus the interaction potential to second order for these states is
\begin{eqnarray}
\lefteqn{\Delta E_{n\ell m,100} = \langle n\ell m|\langle 100| V |100\rangle |n\ell m\rangle} \hspace{10mm}  \nonumber \\&&
+{\sum_{ij}} \frac{\langle n\ell m| \langle 100 | V | a_i \rangle| b_j \rangle \langle b_j |\langle a_i | V | 100\rangle| n\ell m \rangle}{E_{100} +E_{n\ell m} -E_a^i -E_b^j}  
\end{eqnarray}
The summation excludes the states $|a_i\rangle |b_i\rangle = |100\rangle |n\ell m\rangle$ and $|a_i\rangle |b_i\rangle = |n\ell m\rangle |100\rangle$.
At large separation this expression is dominated by the second term and behaves as
\begin{equation}
\Delta E_{n\ell m,100} \approx \frac{C_6(n,\ell ,m)}{R^6}
\end{equation}
where $C_6$ is always negative implying an attractive force unlike the situation for the resonance interaction where the force can be attractive or repulsive.

For p states, only the second order term above applies.  Thus, as suggested by Margenau \& Kestner~(\cite{margenau_kestner}), the interaction energy to second order is the sum of the resonance interaction and the second order dispersive-inductive term such that
\begin{eqnarray}
\lefteqn{\Delta E_{n1m,100} = \langle n1 m|\langle 100| V  |n1 m\rangle |100\rangle} \hspace{10mm}  \nonumber \\&&
+{\sum_{ij}} \frac{\langle n1 m| \langle 100 | V | a_i \rangle| b_j \rangle \langle b_j |\langle a_i | V | 100\rangle| n1 m \rangle}{E_{100} +E_{n1 m} -E_a^i -E_b^j}
\end{eqnarray}  
the summation excluding the degenerate states as above.  For these states however, the first order term dominates due to the resonance interaction.

In order to simplify the infinite sum over all product states of the system in the above second order expressions we employ the first of two approximations suggested by Uns\"old~(\cite{unsold:27}) where, at a fixed separation $R$ between the atoms, the energy denominator is replaced by a constant value $E_p(R)$.  The infinite sum in the above second order expression can then be completed using the closure relation reducing the expression to the simpler form
\begin{eqnarray}
\lefteqn{\Delta E^{(2)}(R)  =   \frac{1}{E_p(R)}[\langle n\ell m| \langle 100| V^2 |100 \rangle | n\ell m \rangle -} \hspace{30mm}\nonumber \\
&&\langle n\ell m| \langle 100| V |100 \rangle |n\ell m \rangle^2 ]
\end{eqnarray}
Furthermore we make the approximation that we may use the value of $E_p$ at infinite separation, $E_p(\infty)$, at all separations $R$.

\begin{table}
\begin{center}
\begin{tabular}{ccc}
\hline
State          &  $C_6$ (a.u.) & $E_p$ (a.u.) \\
\hline
1s1s         & $-6.499027$   & $-0.9232$ \\
1s2s         & $-204.7356$   & $-0.4103$ \\
1s2p$\sigma$ & $-174.1659$   & $-0.4823$ \\
1s2p$\pi$    & $-94.4574$    & $-0.5082$ \\
1s3s$\sigma$ & $-920.477$    & $-0.4498$ \\
1s3p$\sigma$ & $-1117.789$   & $-0.4509$ \\
1s3p$\pi$    & $-632.963$    & $-0.4550$ \\
1s3d$\sigma$ & $-725.402$    & $-0.4466$ \\
1s3d$\pi$    & $-629.676$    & $-0.4574$ \\
1s3d$\delta$ & $-393.8156$   & $-0.4571$ \\
\hline
\end{tabular}
\end{center}
\caption{Implied $E_p$ values for long range H--H interactions computed from the $C_6$ calculations of Stephens \& Dalgarno~(\cite{stephens}). }
\label{tab:eps}
\end{table}

For the first few states of hydrogen we have inferred the value of $E_p$ from previous calculations of the van der Waals coefficient $C_6$ by Stephens \& Dalgarno~(\cite{stephens}), and these are tabulated in Table~\ref{tab:eps}.   For higher states the value of $E_p$ is well approximated by $-4/9$ atomic units, a value obtained by neglecting the contribution to the energy denominator made by virtual states of the excited atom, the second approximation suggested by Uns\"old~(\cite{unsold:55}).  In the case of the long range H--H interaction being considered here it is expected that the value of $E_p$ will converge towards this value for higher lying states, and this is seen to be the case in Table~\ref{tab:eps}, particularly for states with $\sigma$-symmetry which make the largest contribution to the interaction and hence the line broadening.

 In Paper I an $E_p$ value of $-4/9$ was used for all states.  When $E_p$ values from Table~\ref{tab:eps} are used we do not find any significant change to the broadening and consequently the estimate $-4/9$ is used for all higher states.

\subsection{Potential Curves}

The first and second order dispersive-inductive terms, in the context of the Uns\"old approximation, were computed using methods described by Anstee \& O'Mara~(\cite{ao:nad,ao:sp}) and Barklem \& O'Mara~(\cite{bo:pd}).
For p states the total interaction can be obtained by simply adding the second order interactions to the resonance interaction.  

We have made comparisons of a number of our spin-averaged type potential curves with appropriate molecular-type spin dependent curves from the literature.
Our potential curves fulfilled our expectations from previous experience.  That is, they show excellent agreement at long range.  They then have reasonable agreement with molecular curves at ``intermediate'' separations (we define these separations as where the potential starts to deviate from the long range asymptotic expansion behaviour), far better agreement than the multipole expansion results.  At shorter range we see generally poor agreement as the use of perturbation theory breaks down at these separations. For $1$s$2\ell m$ interactions good agreement was observed for $R$ greater than 10--15 $a_0$. It will be shown however that our curves are of acceptable accuracy in the region that is predicted by the model to be important in broadening.

Following Anstee \& O'Mara~(\cite{ao:nad}) the interatomic separations important in the determination of the line broadening cross-sections have been identified by multiplying potentials by a Gaussian "lump" of unit peak amplitude and width of 2 Bohr radii and adding them to original potentials thus amplifying them by up to a factor of two. The line broadening cross-section can then be calculated as a function of the lump position and plotted against the lump position. A corresponding lump appears in this plot which clearly identifies the interatomic separations important in the line broadening.

Using this procedure it was found that the broadening of the 3d state is most sensitive to potentials at intermediate separations, ie. those where the potential curve starts to deviate from the long range behaviour, around 10--30 $a_0$ in this case.  This is the same as has been observed in the broadening of metallic lines by hydrogen collisions (Anstee \& O'Mara~\cite{ao:nad}, Barklem \& O'Mara~\cite{bo:ion1}). Due to the strong resonance interaction with a $R^{-3}$ dependence at long range for 2p states it was found that the broadening is much more sensitive to the long range interaction. Even at a lump position of 60 $a_0$ the cross-section still showed some sensitivity, not yet having converged to the value of 1180 atomic units.  It was also observed that the model appears to be more sensitive to the 1s2p$\pi$ curve than the 1s2p$\sigma$ at intermediate separations, however, we see that at larger separations the broadening is more sensitive to the 1s2p$\sigma$ curve. This behaviour was somewhat unexpected and may be result of the dispersive--inductive interaction being attractive for each of these states while the resonance interaction is attractive for $\sigma$ states and repulsive for $\pi$ states.
In conclusion, it has been shown that the model used here is insensitive to the accuracy of the potentials at small separations where the curves used in this work are known to be inaccurate.

The dependence of the cross-section for the p--d component of H$\beta$ with $E_p$ for the 2p and 4d states was also investigated.  This transition was chosen as the $E_p$ values for the upper state are the most uncertain.  It was seen that the broadening is practically independent of the 2p state $E_p$ value, since this potential is dominated by the first order resonance component which is not dependent on $E_p$.  The dependence on the $E_p$ of the upper state is actually reasonably strong when considering say $E_p$ varying over the range $-0.8$ to $-0.4$, however, it can be safely assumed that the $E_p(\infty)$ values for the 1s4d interactions lie between $-0.457$ and $-0.444$.  Within this range the cross-section was found to deviate by only around 1 per cent.

Our calculations do not include exchange effects.  Our investigations of metallic lines suggest that exchange effects start to become important when $n^*-\ell > 3$ where $n^*$ is the effective principal quantum number which equals $n$ in hydrogen. As shown by Lortet \& Roueff~(\cite{lortet_roueff}), the p--d transition dominates the Balmer lines and exchange effects thus should not be important in the broadening for H$\alpha$, H$\beta$ and H$\gamma$.

\subsection{Validity of Approximations}
\label{sect:valid}

The validity of approximations used in the calculations, is now considered.  All of the assumptions or approximations used in previous work for metallic lines are retained (Anstee \& O'Mara~\cite{ao:nad}).  
We have mentioned the various approximations in the previous discussion of the theory.  The approximations made are the impact approximation (including the binary collision assumption), use of Rayleigh--Schr\"odinger perturbation theory, the classical straight path approximation and the neglect of collision-induced transitions.  In hydrogen lines the impact approximation and neglect of collision-induced transitions may breakdown.  The impact approximation may be suspect in the far line wings due to the fact that the lines are often so broad.  The neglect of collision-induced transitions becomes doubtful as the levels are $\ell$ degenerate.  Hence we will discuss these two approximations.

\begin{table}
\begin{center}
\begin{tabular}{cccc}
\hline
Line      & $\sigma$ & $\alpha$ & $\Delta\lambda_{\mbox{max}}$ \\
          & $(a_0^2)$& &  (\AA) \\
\hline
H$\alpha$  & 1180    & 0.677 & 35.0 \\
H$\beta$   & 2320    & 0.455 & 13.2 \\
H$\gamma$  & 4208    & 0.380 & 7.7 \\
\hline
\end{tabular}
\end{center}
\caption{The broadening characteristics of the p--d component of lower Balmer lines.  The cross-section $\sigma$ is given in atomic units for a collision speed of $10^4$ m s$^{-1}$.  The velocity parameter $\alpha$ gives the velocity dependence assuming $\sigma(v)\propto v^{-\alpha}$.  The approximate maximum detuning for validity of the impact approximation for the self-broadening, computed for $v=14000$ m s$^{-1}$, is given.}
\label{tab:linedata}
\end{table}

The validity of the impact approximation for self-broadening of hydrogen lines was discussed by Lortet \& Roueff~(\cite{lortet_roueff}).  In view of our new calculations we can revisit this analysis, now without the need to split the conditions into resonance and van der Waals parts.  The impact approximation is \emph{strictly} valid when the detuning (in angular frequency units) is far less than the inverse collision duration.  If one considers the detuning for which these quantities are equal, the absolute maximum detuning for which the impact approximation is valid can be estimated by
\begin{equation}
\Delta \lambda_{\mbox{max}} = \frac{\lambda^2}{2\pi c} \frac{v}{\sqrt{\sigma(v)/\pi}}
\end{equation}
where as usual $\sigma(v)$ is the broadening cross-section for collision velocity $v$.  Using the cross-section data which we discuss in the next section (see Table~\ref{tab:linedata}) we have computed $\Delta\lambda_{\mbox{max}}$ for the lower Balmer lines and these are shown in Table~\ref{tab:linedata} for a collision speed of 14000~m~s$^{-1}$.  Such a collision speed corresponds approximately to 5000~K temperature.  For the sun, hydrogen line wings are formed in regions of the atmosphere which are typically hotter than this, and thus $\Delta \lambda_{\mbox{max}}$ is greater as $\bar{v}\propto\sqrt{T}$ and $\sigma\propto T^{(1-\alpha)/2}$ with $0<\alpha<1$, where $\alpha$ is defined in the caption to Table~\ref{tab:linedata}. The impact approximation is only strictly valid when the detuning is far less, say around five times less, than the inverse collision duration. Using this as the criterion the impact approximation is only secure at detunings of less than about 7.0 \AA\ for H$\alpha$ and about 2.6 \AA\ for H$\beta$ in these conditions.

Examining the extent of the solar profiles one sees that the approximation is valid for most of the H$\alpha$ profile but is not valid in the outer wings of the H$\beta$ and H$\gamma$ profiles.
Outside of the limits set in Table~\ref{tab:linedata} one could use methods which are reviewed by Allard \& Kielkopf~(\cite{allard_kielkopf}). However, outside the impact regime collisions at very short range become important where the method we use to calculate the interatomic interaction is no longer valid.

In our calculations it is assumed that the collisions do not cause transitions between nearly degenerate states of the same $n$ but different $\ell$. Such transitions are only likely when the duration of the collision is comparable with the Bohr period for transitions between these nearly degenerate states whose splitting is brought about by the quasistatic ion field. This condition is often termed the Massey criterion. If the collision duration is either much greater (the adiabatic approximation) or much smaller (the sudden approximation) than the Bohr period the probability of a transition occurring is very low. In the present context the collision duration is given by $\bar{b}/v$, where $\bar{b}$ can be estimated from the cross-section data in Table~\ref{tab:linedata} and at unit optical depth in the sun a typical collision speed is about 14000 m s$^{-1}$. The appropriate Bohr periods for H$\alpha$, H$\beta$ and H$\gamma$ can be estimated from linear Stark shift parameters for hydrogen (for example see Condon \& Shortley~\cite{condon_shortley}) and an estimate of the quasistatic ion field at unit optical depth in the sun. A comparison of the collision durations and Bohr periods show that the Bohr period is about 400 times greater for H$\alpha$, 100 times greater for H$\beta$, and 50 times greater for H$\gamma$, than the collision duration at unit optical depth in the solar photosphere. These results indicate that above unit optical depth in the sun, the sudden approximation is valid and that for these lines $\ell$-changing collisions can be neglected. Due to the increasing Stark effect $\ell$-changing collisions may become important for the higher Balmer lines. Due to the lower quasistatic ion field in metal deficient stars $\ell$-changing collisions will be even more unlikely.

\section{Results}

\begin{figure}
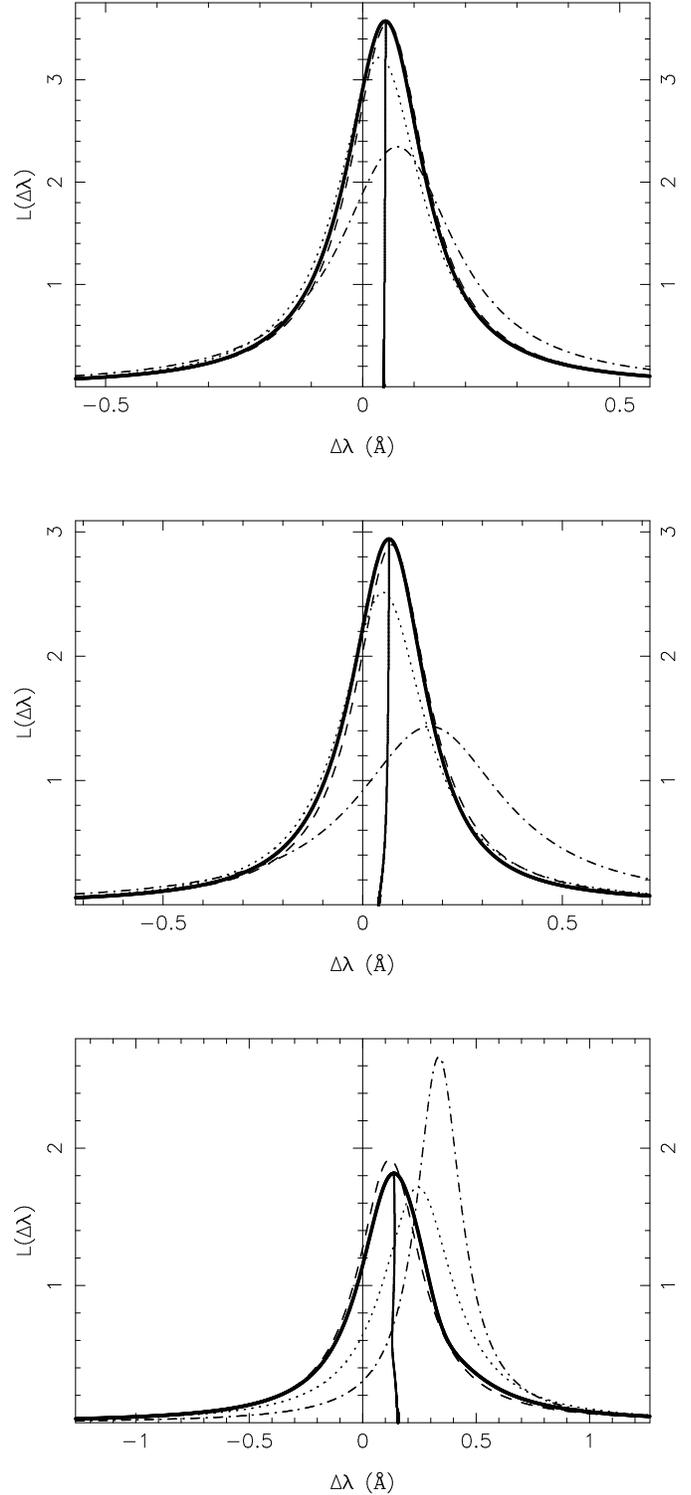

\resizebox{\hsize}{!}{\rotatebox{270}{\includegraphics{h2311.f2a}}}

\vspace{8mm}

\resizebox{\hsize}{!}{\rotatebox{270}{\includegraphics{h2311.f2b}}}

\vspace{8mm}

\resizebox{\hsize}{!}{\rotatebox{270}{\includegraphics{h2311.f2c}}}

\caption{Hydrogen broadened profiles for H$\alpha$ (top), H$\beta$ (middle) and H$\gamma$ (bottom) at 8000 K and $10^{18}$ perturbers per cubic cm.  Shown are the complete profile from Eq.~(\ref{eq:lineshape}) (full), the p--d component (dashed), the s--p component (dotted) and the p--s component (dot-dash).  The full vertical line is the line bisector of the full line profile. All profiles are area normalised. Note the different scales, and that due to the different central wavelengths of the lines the above widths should not be directly compared with each other.}
\label{fig:profiles}
\end{figure}

In the absence of ions and electrons, profiles, for a given temperature and hydrogen atom perturber density, have been computed for H$\alpha$, H$\beta$ and H$\gamma$ using overlapping line theory. Not only are such profiles valuable in examining the contributions made by the component lines to each profile but these profiles represent a limiting real situation in the spectra of cool stars as the metallicity is reduced to zero. In Fig.~\ref{fig:profiles} the profiles obtained from overlapping line theory are plotted along with profiles for the three component lines which contribute to the overlapping line profile.  These three components, each weighted by the appropriate dipole matrix element, sum to give the profile.

It was pointed out by Lortet \& Roueff~(\cite{lortet_roueff}) that the p--d component of the Balmer lines is by far the strongest.  This is clearly seen in Fig.~\ref{fig:profiles}, where the total profile and p--d component profile are very similar.
In Fig.~\ref{fig:profiles} we see that each component (all Lorentzian and therefore symmetric) has a different predicted pressure induced line shift.  This leads to a very small predicted asymmetry in the total predicted line profile.  However, we should comment that we expect that the shift calculations of our method are less reliable than those for widths, as shifts are perhaps more dependent on strong collisions and hence the short range interaction potential (Anstee \& O'Mara~\cite{ao:nad}). 

One also sees a marked difference in the relative width of the p--s component of H$\gamma$ compared to the other two lines.  This component is relatively narrow, whereas in the other lines it is the broadest component.  We expect the broadening of this component to be seriously overestimated in both H$\beta$ and H$\gamma$ due to neglect of exchange effects.  Fortunately however, this component has almost negligible effect on the overall line profile.

\subsection{Temperature Dependence}

Previous theories of resonance broadening predict a line width which is independent of temperature. This is a result of the interaction decreasing with increasing separation like $R^{-3}$ which leads to a cross-section which is inversely proportional to the collision speed. For the 2p state (resonance interactions only) we obtain essentially the same result but for more excited p states we observe a temperature dependence which increases with increasing excitation. This temperature dependence becomes quite significant for the 5p state. This difference is a result of increasing departure in our calculations from an $R^{-3}$ dependence of the interaction on the interatomic separation indicating that the multipole expansion used in previous calculations is only strictly valid for the 2p state.

When one introduces the dispersive-inductive interactions the self-broadening is found to be no longer temperature independent even for the low lying states.  This result is of astrophysical importance as we will discuss later.

\subsection{Comparison with Ali \& Griem}

\begin{figure}
\resizebox{\hsize}{!}{\rotatebox{90}{\includegraphics{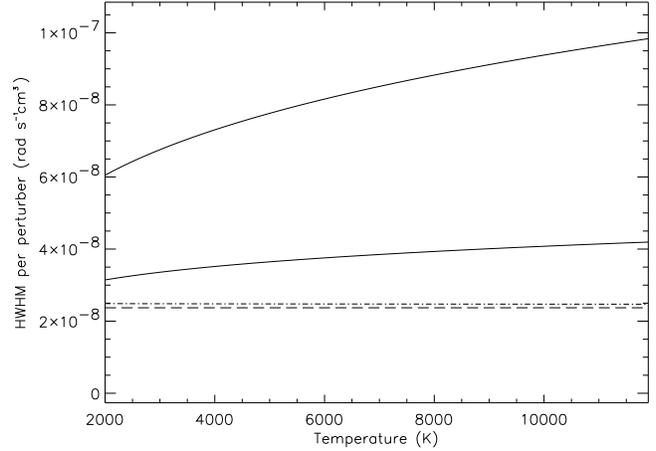}}}
\caption{Comparison of the line width (HWHM) per perturber with temperature computed in this work for the dominating 2p--3d component of H$\alpha$ (lower full) and 2p--4d component of H$\beta$ (upper full) with that of the resonance broadening theory of Ali \& Griem (dashed) for the 2p state, and our calculation of the resonance broadening (dot-dash) for this state.}
\label{fig:widths}
\end{figure}

Fig.~\ref{fig:widths} compares line widths from our treatment of self-broadening with those of Ali \& Griem~(\cite[corrected]{ali_griem:errata}), which include only the resonance broadening of the lower 2p state, for H$\alpha$ and H$\beta$ as a function of temperature.  We find that our results are in quite good agreement with the Ali \& Griem~(\cite{ali_griem:errata}) theory when we only consider the resonance interactions as they did.  However, we find that the effect of the dispersive--inductive interaction of other states involved in the transition is quite substantial, particularly that resulting from the d state of the upper level in Balmer lines.
The dispersive contribution relative to the resonance contribution for H$\beta$ is greater than for H$\alpha$ and this is reflected in the stronger temperature dependence of the line width.

\begin{figure}
\resizebox{\hsize}{!}{\rotatebox{90}{\includegraphics{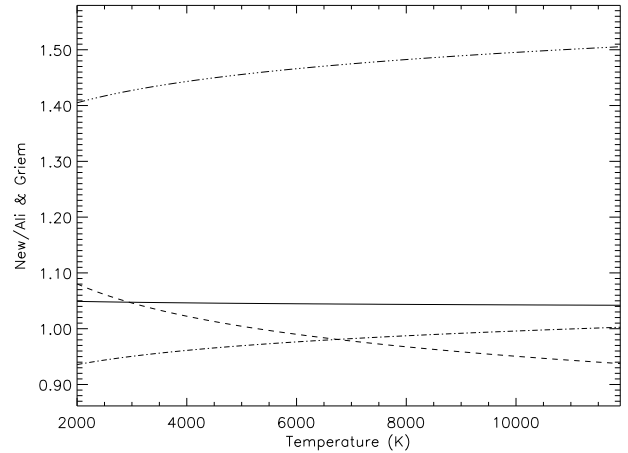}}}
\caption{Comparison showing the ratio of our results for resonance broadening and the Ali \& Griem~(\cite{ali_griem:errata}) theory.  The full, dashed, dot-dash and dot-dot-dot-dash lines correspond to the 2p, 3p, 4p and 5p levels respectively.}
\label{fig:ag_ours}
\end{figure}

Fig.~\ref{fig:ag_ours} shows the ratio, as a function of temperature, of the line widths resulting from our treatment of resonance broadening of the 2p, 3p, 4p, and 5p states with those of Ali \& Griem~(\cite{ali_griem:errata}).  The difference can be attributed to a failure of the multipole expansion of the electrostatic interaction between the two atoms which is used by Ali \& Griem~(\cite{ali_griem:errata}).  Lortet \& Roueff~(\cite[Fig.~3]{lortet_roueff}) show calculations that suggest that the multipole expansion should breakdown for all p states but the 2p state.  Our calculations suggest that the break down is seen for these states but is not severe until the 5p state for the collision speeds of interest here.

\subsection{Approximation by p--d Component in Balmer lines}

Using overlapping line theory, grids of profiles which result from self-broadening alone have been computed for a range of temperatures and hydrogen atom number densities from which one can interpolate the appropriate profile for a given set of physical conditions. However, we have already seen that these complete profiles obtained from overlapping line theory are very closely approximated by the p--d component of the relevant Balmer line. When applied to synthetic Balmer lines in the solar spectrum for H$\gamma$ the maximum difference is less than 0.2 percent of the continuum flux, less for H$\beta$, and even less for H$\alpha$. Errors resulting from employing the p--d approximation in the interpretation of real stellar spectra will lie in the noise associated with the observational data. 

A possible objection to the p--d approximation is that it has been developed in the limit of a zero quasi-static ion field. The quasi-static ion field destroys spherical symmetry leading to $\ell$ no longer being a good quantum number so that it is no longer strictly possible to talk about a p--d transition. This will certainly be the case in hot stars where the ions are protons produced by the almost complete ionisation of hydrogen leading to a strong quasi-static ion field. However, our interest is in cool stars where the ions are produced by thermal ionisation of metals.  The quasistatic ion field is proportional to $N_i^{2/3}$ where $N_i$ is the ion number density which in cool stars is smaller than that in hot stars, where the ions are largely protons, by four orders of magnitude for a star with solar composition and by perhaps six orders of magnitude for cool stars of low metallicity. Therefore in this work we are working in the limit of very weak quasistatic ion fields. Under these circumstances the p and d states are only very weakly mixed by the weak quasistatic ion field with states of other $\ell$ with the same $n$. Thus in the limit of very weak quasistatic ion fields $\ell$ is an almost good quantum number and the p--d approximation is acceptable.

The p--d approximation neglects the line shift and asymmetry predicted using overlapping line theory. However, when overlapping line theory and the p--d approximation are used in the synthesis of Balmer lines in the solar spectrum the shifts and asymmetries predicted by overlapping line theory are not detectable due to the effects of Stark broadening and the profiles are in good agreement with those predicted using the p--d approximation. Due to the reduced ion/electron density, synthetic spectra for very cool stars (T $<$ 4500 K) show some evidence of shift and asymmetry when overlapping line theory is used. However, the synthesis of the spectrum of very cool stars is complicated by impact broadening due to molecular hydrogen which is not included in our calculations.

Although the p--d approximation does not significantly reduce computing time it does permit self-broadening data to be presented in a way which is much more efficient than the publication of grids of line profiles. The data relevant to the application of the p--d approximation to the first three Balmer lines are presented in Table~\ref{tab:linedata}. Data in the form of grids can be obtained from the authors. In spite of the advantages of the p--d approximation grids have been used in all calculations in this paper and in Paper I.

\section{Comparison of Broadening Mechanisms}

We now compare the relative strengths of broadening mechanisms in the wings of lower Balmer lines through a model solar atmosphere.  As the profiles are not necessarily of the same shape, the best way to do this is to compare the depth of the normalised profile at some suitable detuning from line centre.

\begin{figure*}
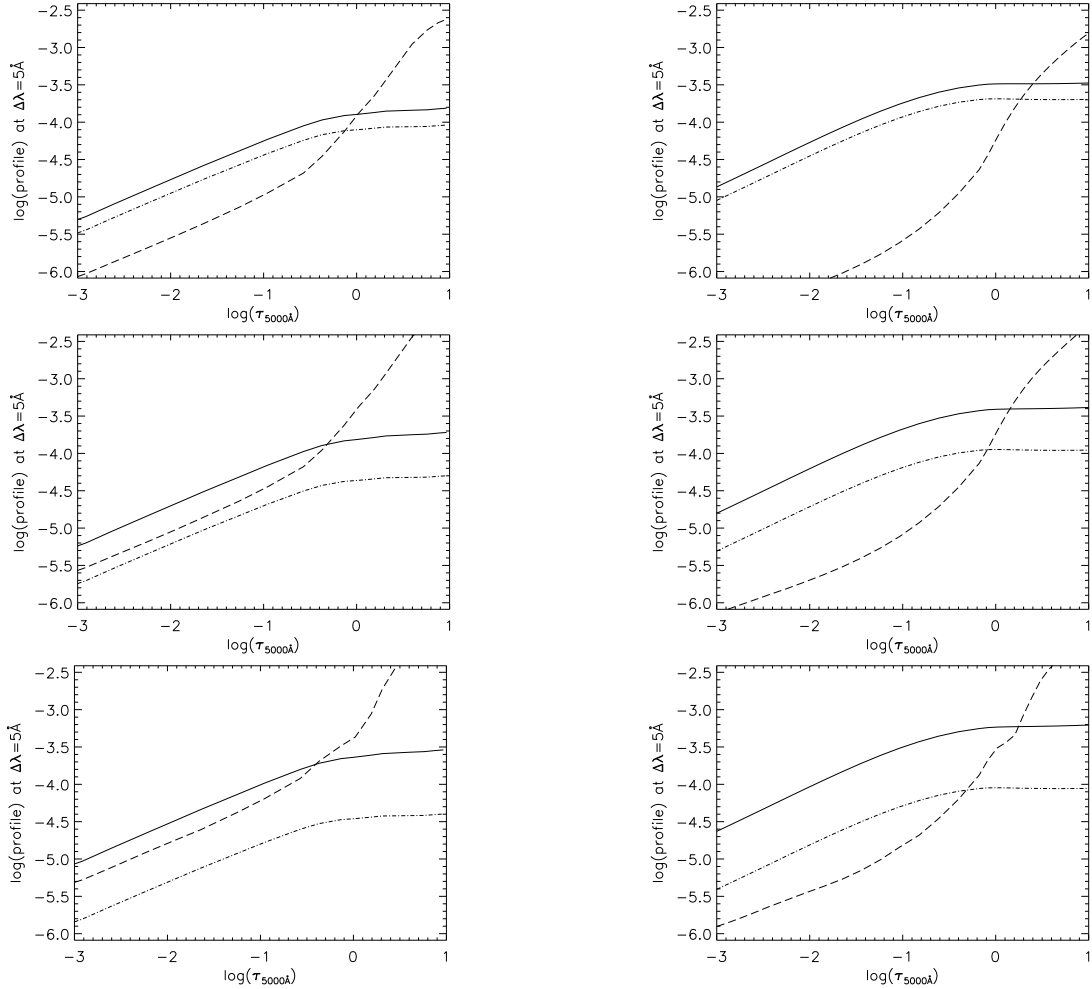

\begin{center}
\tabcolsep=10mm
\begin{tabular}{cc}

\vspace{-5mm}

\resizebox{65mm}{!}{\rotatebox{90}{\includegraphics{h2311.f5a}}} & \resizebox{65mm}{!}{\rotatebox{90}{\includegraphics{h2311.f5d}}} \\

\vspace{-5mm}

\resizebox{65mm}{!}{\rotatebox{90}{\includegraphics{h2311.f5b}}} & \resizebox{65mm}{!}{\rotatebox{90}{\includegraphics{h2311.f5e}}} \\

\resizebox{65mm}{!}{\rotatebox{90}{\includegraphics{h2311.f5c}}} & \resizebox{65mm}{!}{\rotatebox{90}{\includegraphics{h2311.f5f}}}

\end{tabular}
\end{center}
\caption{The broadening mechanisms through the model solar atmosphere of Holweger \& M\"uller~(\cite{holmul}) (left) and a MARCS (Asplund~et~al.~\cite{osmarcs}) model atmosphere of a metal poor [Fe/H]$ = -2.0$ star of solar temperature and gravity (right) for H$\alpha$ (top), H$\beta$ (middle) and H$\gamma$ (bottom).  The lines plot the depth of the line profile for our self--broadening theory (full), the Ali \& Griem~(\cite{ali_griem:errata}) resonance broadening theory (dot-dash) and for the Stehl\'e~(\cite{stehle:94}) Stark broadening theory (dash), at 5\AA\ detuning from the line centre.  Note that the Stark profiles are folded with Doppler profiles, however, Doppler profiles make negligible contribution at this detuning for these temperatures.  }
\label{fig:mechs}
\end{figure*}

We plot the profile depth of each broadening mechanism profile at 5 \AA\ detuning through the Holweger \& M\"uller~(\cite{holmul}) model solar atmosphere in Fig.~\ref{fig:mechs}.  When one considers that the wings of the lines are formed in the region around $\log \tau_{5000 \mathrm{\AA}}=0$, we clearly see that the new theory makes a significant difference when compared with Ali \& Griem's theory.  In this region, rather than being weaker than the Stark broadening contribution, the contribution of self-broadening is now often comparable.

In cool metal poor stars, the electrons and ions are out-numbered by hydrogen atoms by an even greater number than in stars around solar metallicity.  Thus self-broadening in hydrogen lines becomes even more important.  Fig.~\ref{fig:mechs} also shows similar plots to those shown for the solar model, for a MARCS model (Asplund~et~al.~\cite{osmarcs}) with solar temperature ($T=5770$~K) and surface gravity ($\log g=4.44$) but [Fe/H]$ = -2.0$.  Here we see clearly that the new theory has a significant effect on the contribution to broadening in the line forming region.  At great depth the Stark broadening always dominates due to the higher ion/electron density while self-broadening (using our theory) dominates above an optical depth of 0.1 for solar composition and above an optical depth of 1 in the metal deficient case. 

\section{Synthetic Stellar Spectra}

The computation of synthetic stellar spectra requires the convolution of all broadenings. We convolve our self-broadened profiles with appropriate Stark profiles from Stehl\'e~(\cite{stehle:94}) which are provided preconvolved with the Doppler profiles.  The profiles are then further convolved with profiles for radiative and helium collision broadening.  In these calculations we approximate the convolution in the far wings by adding the profiles (Stark, self-broadening, radiative and helium broadening) together, following the Kurucz~(\cite{kurucz:cds}) codes. 

This procedure can be justified, for cool stars, in terms of the p--d approximation which we know to be valid in the weak quasi-static ion field limit which we know to exist in such stars. In the p--d approximation in the absence of ions the line is well represented by the p--d component alone which will have a Lorentz profile due to impact broadening by electrons and hydrogen atoms. In the presence of a given weak quasi-static ion field this profile will be Stark shifted by an amount dictated by the first order Stark shift of the p--d component. The final profile can then be found by integrating this profile over the Holtsmark distribution of quasi-static ion fields which in the weak field limit will be well approximated by a Lorentzian with a width which is the sum of the electron impact width and self-broadening width somewhat enhanced by the smearing effect of the quasi-static ion field. Using this as a guide an alternative procedure is to calculate the profile in the absence of self-broadening using, for example, the profiles of Stehl\'e~(\cite{stehle:94}). In the weak quasi-static ion field limit these profiles should be well approximated by a Lorentzian (for example Stehl\'e~\cite{stehle:96}) with the full impact width containing all line components (but dominated by the p--d component) somewhat enhanced by the smearing effect of the weak quasi-static ion field. As the profile is Lorentzian in the wings the absorption will be proportional to this enhanced impact width. The profile of the line in the absence of ions and electrons produced by self-broadening will also be Lorentzian with a depth in the wings proportional to the self-broadening impact width which again contains the effect of all components but dominated by the p--d component. Thus all three sources of broadening can be represented by a Lorentzian with a width which is simply the sum of the widths of the two profiles or equivalently in the line wings by simply adding the profiles. The effect of radiative broadening and broadening by helium collisions can be included in the same way.

Test calculations show this procedure for the convolution to be an excellent approximation for the cases considered here. For example in the solar synthetic profile, no difference can be seen between the profile computed in this way and that computed with a complete numerical convolution.  The approximation gradually becomes worse in cool stars, and starts to break down in models of effective temperatures around 4000 K, as the lines are no longer strong enough for this approximation to be valid.  

We use the spectral synthesis code of Piskunov~(\cite{piskunov:synth}) for the radiative transfer, which assumes LTE.  Radiative broadening and collisional broadening by helium are included in all calculations, though are found to be negligible in most conditions.

\subsection{The Impact of the Self-Broadening Calculations on Line Profiles}

The most interesting question, is how much difference the theory makes to predicted stellar line profiles when compared to the Ali \& Griem~(\cite{ali_griem:errata}) theory, and thus the commonly used Kurucz~(\cite{kurucz:cds})/Peterson~(\cite{peterson}) codes.

\begin{figure}
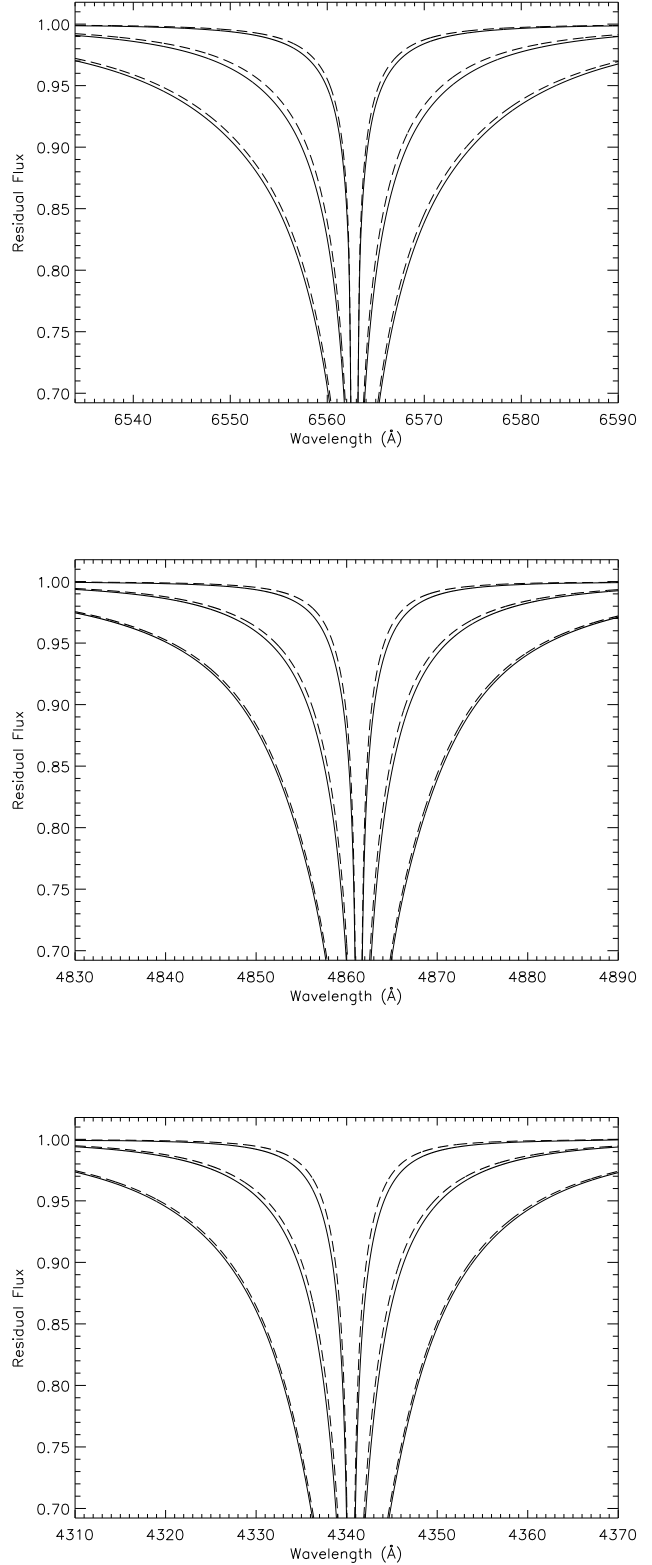

\resizebox{\hsize}{!}{\rotatebox{90}{\includegraphics{h2311.f6a}}}

\vspace{9mm}

\resizebox{\hsize}{!}{\rotatebox{90}{\includegraphics{h2311.f6b}}}

\vspace{9mm}

\resizebox{\hsize}{!}{\rotatebox{90}{\includegraphics{h2311.f6c}}}

\caption{Synthetic flux profiles for H$\alpha$ (top), H$\beta$ (middle) and H$\gamma$ (bottom) for MARCS models of $T_{\mathrm{eff}}$ = 5000, 6000 and 7000 K (top to bottom) for solar gravity and metallicity.  The full lines use our line broadening theory and dashed lines use Ali \& Griem's resonance broadening theory for the hydrogen broadening.}
\label{fig:stellar_profiles}
\end{figure}

\begin{table}
\begin{center}
\begin{tabular}{cccc}
\hline
$T_{\mathrm{eff}}$   & H$\alpha$  & H$\beta$ & H$\gamma$ \\
(K) & & & \\
\hline
5000                 & 12.7       & 15.6     & 17.6      \\
6000                 & 10.3       & 7.7      & 7.6       \\
7000                 & 4.6        & 2.5      & 2.3       \\
\hline
\end{tabular}
\end{center}
\caption{Percentage increases in equivalent width using our self-broadening theory compared with Ali \& Griem~(\cite{ali_griem:errata}) for the synthetic lower Balmer line profiles computed for MARCS models of various effective temperature, with solar gravity and metallicity.}
\label{tab:ew_increases}
\end{table}

Fig.~\ref{fig:stellar_profiles} shows computed line profiles for MARCS models~(Asplund~et~al.~\cite{osmarcs}) for a range of effective temperatures at solar gravity and metallicity, using both our theory and the Ali \& Griem~(\cite{ali_griem:errata}) resonance broadening theory.  Table~\ref{tab:ew_increases} shows the increase in the equivalent width brought about by our self-broadening theory.
Although Fig.~\ref{fig:widths} indicates the effect of the new theory on the self-broadening is larger in H$\beta$ than H$\alpha$ this is not seen in the synthetic stellar spectra in Fig.~\ref{fig:stellar_profiles} due to the fact that the Stark broadening profile widths are increased by an even greater amount, as shown in Fig.~\ref{fig:mechs}.

The decline in the difference between the two theories with increasing temperature is due to the increase in the Stark broadening resulting from ionisation of hydrogen as temperature increases. For stars earlier than F type the self-broadening will become irrelevant as it will be completely overwhelmed by Stark broadening.

\subsection{Predicted Impact on Effective Temperature Determinations}

\begin{figure}
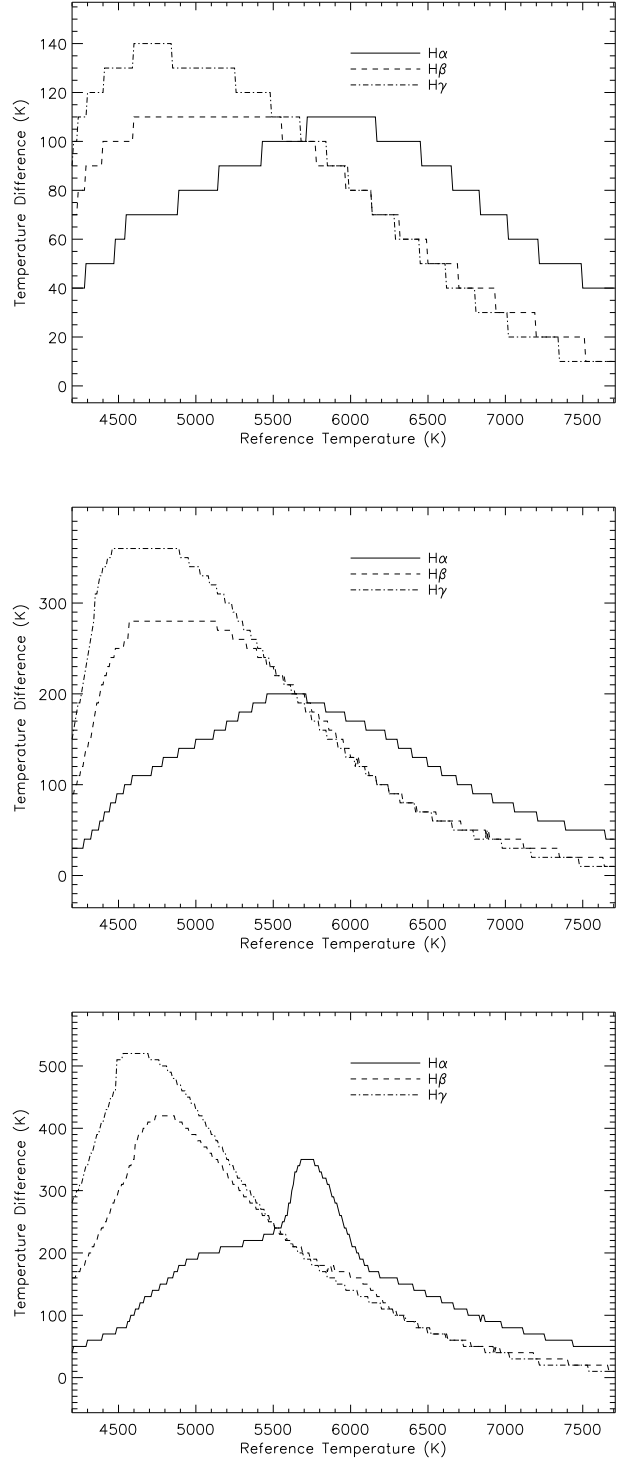


\resizebox{\hsize}{!}{\rotatebox{90}{\includegraphics{h2311.f7a}}} 

\vspace{2mm}

\resizebox{\hsize}{!}{\rotatebox{90}{\includegraphics{h2311.f7b}}} 

\vspace{2mm}

\resizebox{\hsize}{!}{\rotatebox{90}{\includegraphics{h2311.f7c}}}

\caption{The predicted difference in effective temperature determinations from our new calculations, and calculations using the resonance broadening theory of Ali \& Griem~(\cite{ali_griem:errata}) for H$\alpha$, H$\beta$ and H$\gamma$.  Plots are shown for solar metallicity (top), 1/10 solar metallicity (middle) and 1/100 solar metallicity (bottom), in both cases for solar surface gravity.  The ``reference temperature'' is that which would be found using our broadening theory.  The plot then predicts how much higher the effective temperature derived using the Ali \& Griem~(\cite{ali_griem:errata}) theory is expected to be.}
\label{fig:teffdiffs}
\end{figure}

In Paper I we made preliminary estimates of the effect that the new calculations would have on the determination of effective temperatures.  Here we present an extended analysis including H$\gamma$ calculations and covering higher effective temperatures.  Also included are new estimates for [Fe/H]${}=-1.0$.  The low metallicity calculations have also been redone on a finer model grid.

We computed a grid of MARCS models (Asplund~et~al.~\cite{osmarcs}) over a range of temperatures, with solar gravity, for metallicities of [Fe/H]${}=0.0, -1.0$ and $-$2.0. We used the grid to estimate the difference in $T_{\mathrm{eff}}$ determined from our theory and the Ali \& Griem~(\cite{ali_griem:errata}) theory. For each model we computed synthetic profiles as described above using both our theory and the Ali \& Griem~(\cite{ali_griem:errata}) theory.  For each profile resulting from our theory we then found the best matching profile (in the line wings) using the Ali \& Griem~(\cite{ali_griem:errata}) theory, and recorded the temperature difference between the models used to generate the two profiles.  The results, plotted in Fig.~\ref{fig:teffdiffs}, indicate that the new line broadening calculations lead to a significant lowering of the derived effective temperature.

We see that the new results for H$\gamma$ are extremely similar to those for H$\beta$.  This is explained by Fig.~\ref{fig:mechs} where we see the relative contribution of Stark and self-broadening through the atmospheres are relatively similar in these two lines.

As pointed out in Paper I, the peak temperature ``error'' and the difference in location of the peak for H$\alpha$ from the other two lines is of interest. Synthetic profiles obtained using our theory are always stronger than those obtained using Ali \& Griem~(\cite{ali_griem:errata}) theory. In Ali \& Griem~(\cite{ali_griem:errata}) theory the H-atom broadening is resonance broadening only and is therefore temperature independent while in our theory the dispersive-inductive contribution leads to an increase with temperature. At low $T_{\mathrm{eff}}$, H-atom broadening makes its greatest contribution and as $T_{\mathrm{eff}}$ is raised the temperature ``error'' increases because of the growth in the H-atom broadening in our theory. Eventually Stark broadening begins to dominate accounting for the peak followed by a decline as Stark broadening becomes more and more dominant as $T_{\mathrm{eff}}$ increases. As Stark broadening in H$\beta$ and H$\gamma$ is greater than in H$\alpha$ the peak occurs at a lower $T_{\mathrm{eff}}$. The higher peak temperature ``error'' for metal poor stars reflects the higher temperature required to increase the ion/electron density sufficiently.  This can be tested observationally.  In agreement with this result Gardiner~et~al.~(\cite{gardiner}, Fig.~9), with mixing length parameter $\alpha=1.25$ and using the Ali \& Griem~(\cite{ali_griem:errata}) theory, found that $T_{\mathrm{eff}}$ obtained from H$\alpha$ is larger than for H$\beta$ at $T_{\mathrm{eff}}$ around 6000--7000K while the situation is reversed for stars with a lower $T_{\mathrm{eff}}$ although admittedly there is only a small sample of stars in this domain. It is perhaps significant to note that Castelli~et~al.~(\cite{castelli}) find, using Ali \& Griem~(\cite{ali_griem:errata}) theory and the solar KOVER model, that $T_{\mathrm{eff}}$ has to be raised by 100--150K (consistent with the peak of 120K for H$\alpha$ in Fig.~\ref{fig:teffdiffs}) in order to fit the observed solar profiles.

\section{Comparison with Solar Observations}

The comparison with observed solar spectra requires the use of a photospheric model.  In this work we use one-dimensional plane-parallel models which are freely available.  Namely these are the semi-empirical solar model of Holweger \& M\"uller~(\cite{holmul}) hereafter HOLMUL, a MARCS theoretical solar model~(Asplund~et~al.~\cite{osmarcs}), and two Kurucz theoretical solar models~(Kurucz~\cite{kurucz:cds}; Castelli~et~al.~\cite{castelli}).  The two types of Kurucz models used, that with and that without convective overshooting, are hereafter KOVER and KNOVER models respectively.   For both MARCS and Kurucz models we use the default mixing length parameters $\alpha$, namely $1.25$ for Kurucz and $1.5$ for MARCS.  We retain the default structure parameters $y$, namely $1/2$ for Kurucz and $3/(4\pi^2)$ for MARCS.

It is expected that MARCS and KNOVER models are quite similar as they are both based on essentially the same physics and ``standard'' mixing length convection theory although with different parameters.  For the solar models used here, the computed Balmer line profiles of MARCS and KNOVER were in excellent agreement.  Hence below we will only discuss the KNOVER model.  However we caution that this agreement may not extend to other stellar parameters. In the KOVER models Kurucz has introduced ``approximate overshooting'' to the convection treatment.  The approximate overshooting assumes ``the centre of a bubble stops at the top of the convection zone so that there is convective flux one bubble radius above the convection zone.  That flux is found by computing the convective flux in the normal way and then smoothing it over a bubble diameter'' (Kurucz~\cite{kurucz:iau92}).

The purpose of this comparison is to test the broadening theory, not the models or convection treatments. The validity of the theory is tested by comparison of Balmer line results with those of other model predictions such as limb-darkening curves.  This situation is clearly not ideal due to uncertainties in the models and the particular sensitivity of Balmer lines to deep layers and convection treatment.  However, lack of laboratory data makes this our best option at present.  3D convective models will be investigated in future.

\subsection{Profile Comparisons}

Limb-darkening curves are a powerful test of solar models. On this basis alone HOLMUL is the preferred model as it reproduces limb-darkening curves
better than either KOVER, KNOVER or MARCS (Blackwell~et~al.~\cite{blackwell}; Castelli~et~al.~\cite{castelli}). However Castelli~et~al.~(\cite{castelli}) found that in spite of KNOVER being unable to reproduce limb-darkening curves as well as KOVER it produces a better fit to hydrogen line profiles when Ali \& Griem~(\cite{ali_griem:errata}) theory is used.

\begin{figure*}
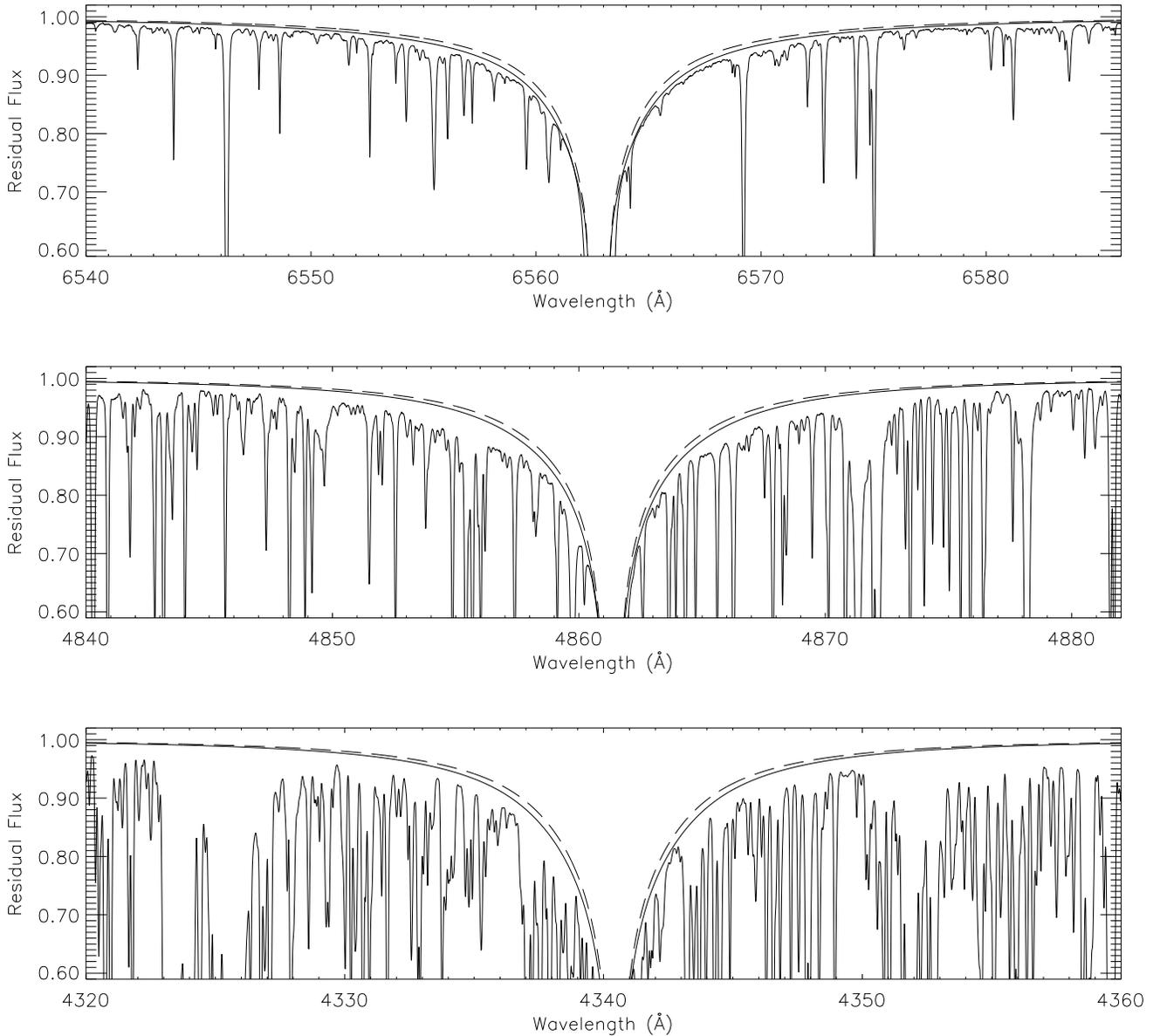

\resizebox{\hsize}{!}{\rotatebox{90}{\includegraphics{h2311.f8a}}}
\resizebox{\hsize}{!}{\rotatebox{90}{\includegraphics{h2311.f8b}}}
\resizebox{\hsize}{!}{\rotatebox{90}{\includegraphics{h2311.f8c}}}
\caption{Comparisons of synthetic flux profiles with observations (NSO/Kitt Peak FTS data) for the sun for H$\alpha$ (top), H$\beta$ (middle) and H$\gamma$ (bottom).  All models lines use the HOLMUL model.  The full line uses our self-broadening theory while the dashed line uses the Ali \& Griem~(\cite{ali_griem:errata}) theory.}
\label{fig:solarcomp2}
\end{figure*}

\begin{figure*}
\resizebox{\hsize}{!}{\rotatebox{90}{\includegraphics{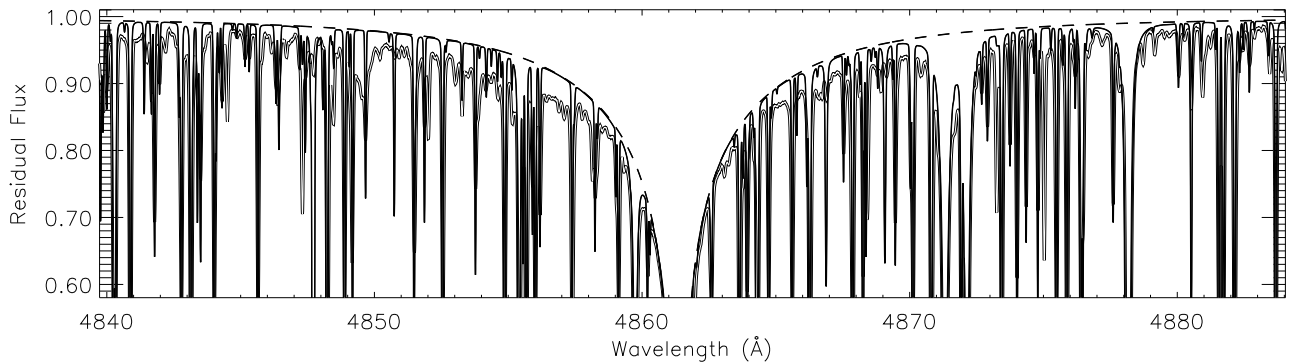}}}
\caption{Comparison of synthetic flux profiles with observations (double line -- NSO/Kitt Peak FTS data) for the sun for H$\beta$ both with (full line) and without (dashed line) blending lines from VALD, both employing the HOLMUL model and our self-broadening theory.  Macroturbulence of 1.6 km s$^{-1}$ and rotational velocity of $v \sin i = 1.8$ km s$^{-1}$ are used.}
\label{fig:solarcomp_beta_wblends}
\end{figure*}

As HOLMUL is the preferred model on the basis of limb-darkening data and its ability to reproduce the behaviour of a large sample of strong metallic lines, computed synthetic profiles for HOLMUL using both our self-broadening theory and Ali \& Griem~(\cite{ali_griem:errata}) theory are compared with the observed solar flux spectrum of Kurucz~et~al.~(\cite{kurucz_atlas}, NSO/Kitt peak FTS data) in Fig.~\ref{fig:solarcomp2}.  We do not adjust the H$\beta$ continuum here as in Paper~I. It is seen that for all three lines our broadening theory reduces the discrepancy with observation but the remaining discrepancy is still significant. As line blending is significant, particularly in H$\beta$ and H$\gamma$ profiles, computations were performed which included all available lines from VALD, the Vienna Atomic Line Database (Kupka~et~al.~\cite{vald}) for all three Balmer lines. The predicted residual fluxes with and
without blending were found to be in good agreement in the windows between the blending lines, as shown for H$\beta$ in Fig.~\ref{fig:solarcomp_beta_wblends}.  However the inclusion of blending lines does not change the conclusion that the synthetic profiles are too weak to match the observations. As there are many blending lines without data or unidentified, particularly for H$\beta$ and H$\gamma$, there is some element of uncertainty in this conclusion.

\begin{figure*}
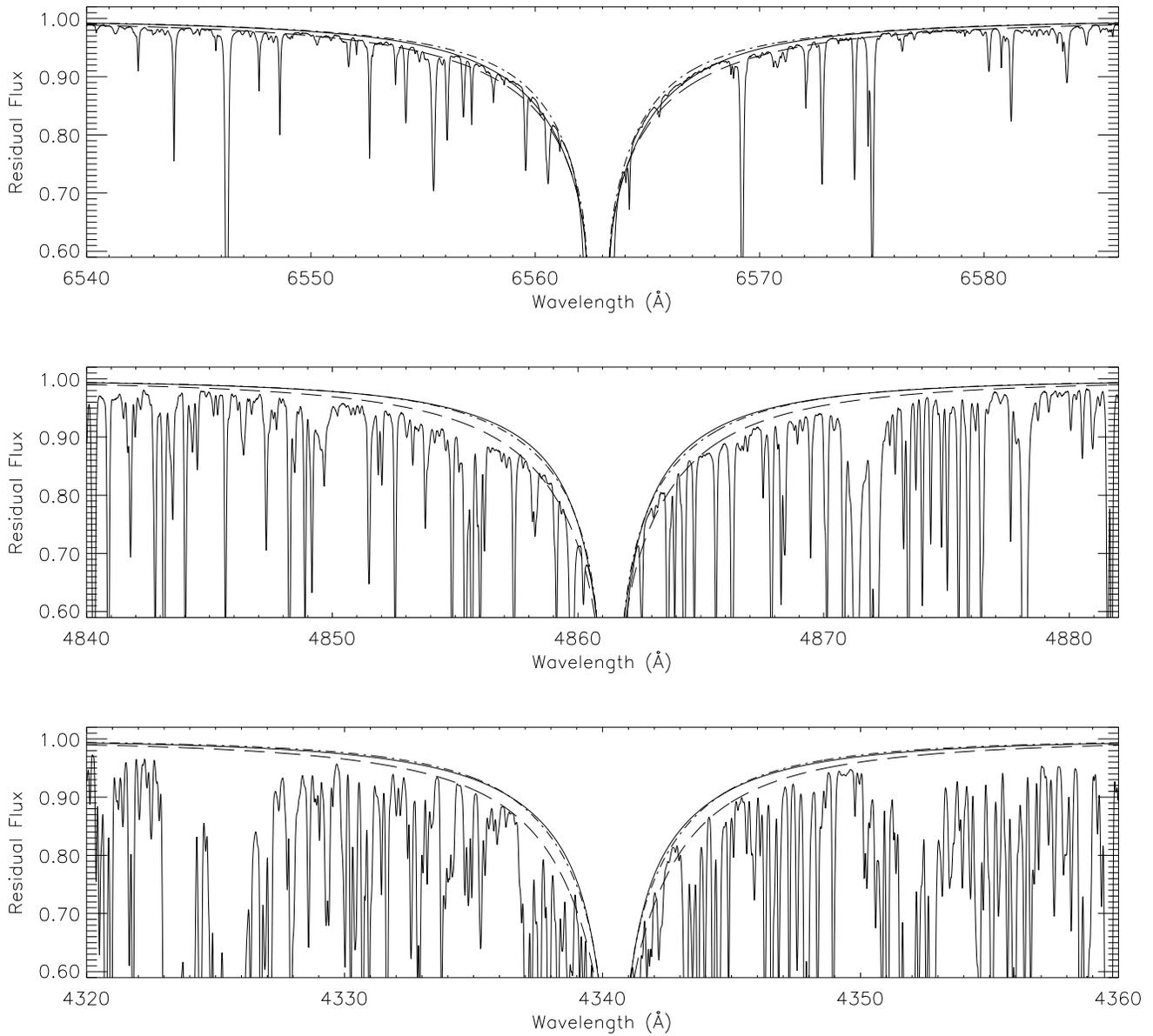

\resizebox{\hsize}{!}{\rotatebox{90}{\includegraphics{h2311.f10a}}}
\resizebox{\hsize}{!}{\rotatebox{90}{\includegraphics{h2311.f10b}}}
\resizebox{\hsize}{!}{\rotatebox{90}{\includegraphics{h2311.f10c}}}
\caption{Comparisons of synthetic flux profiles with observations (NSO/Kitt Peak FTS data) for the sun for H$\alpha$ (top), H$\beta$ (middle) and H$\gamma$ (bottom).  Full and dashed lines use the KOVER and KNOVER models respectively, and the dot--dashed lines use the HOLMUL model.}
\label{fig:solarcomp}
\end{figure*}

In Fig.~\ref{fig:solarcomp} predicted Balmer line profiles using our theory for HOLMUL, KOVER and KNOVER solar models are compared with the observed spectrum. KNOVER predicts profiles for all lines that are generally too strong. If blending lines are included the discrepancy is even greater so KNOVER is now a model which fits neither the limb-darkening nor the Balmer line profiles and is therefore strongly ruled out by our line-broadening theory. For H$\alpha$ and H$\beta$ the synthetic profiles obtained using the KOVER and HOLMUL models are in good agreement with each other but are insufficiently strong to match the observed profiles. In the outer parts of these profiles the discrepancy may be due the failure of the impact approximation (see Table~\ref{tab:linedata}) an inadequate temperature structure or both. In the far wings of H$\alpha$ the synthetic profiles obtained using the KOVER and HOLMUL models are again too weak. Within 5~\AA\ of line centre the profile predicted by the KOVER model is too strong which weakly favours the HOLMUL model. The observed core of the line, within 0.7~\AA\ of line centre, the observed profile is much stronger than any of the synthetic profiles. This part of the line is formed in the low chromosphere which is not included in our synthetic modelling.

In summary our self-broadening theory is superior to the Ali \& Griem~(\cite{ali_griem:errata}) theory because it reduces the discrepancy between the observed and computed Balmer line profiles when the preferred HOLMUL model is used and leads to the KNOVER model being discarded thus resolving the dilemma posed by a model which provides the best match to the Balmer line profiles but fails to match limb-darkening curves when the Ali \& Griem~(\cite{ali_griem:errata}) theory is used. In spite of these successes significant discrepancies remain between theory and observation. However the behaviour of the KNOVER model suggests that it may be possible to construct a model with a temperature structure somewhere between the HOLMUL and KNOVER models which provides the best simultaneous match to the limb darkening curves and the H$\alpha$ profile where the validity of the impact approximation is not an issue. The impact approximation is an important issue for H$\beta$ and H$\gamma$. Fitting of the profiles of these lines should be confined to the detunings indicated in Table~\ref{tab:linedata} and even then with some caution as these detunings correspond to the extreme limit of validity of the impact approximation.

\section{Concluding Remarks}

We have presented a theory of self-broadening of hydrogen lines, which includes both resonance and dispersive--inductive interactions. The theory was used to synthesise Balmer lines in cool stars and shown to make a considerable difference compared to the commonly used Ali \& Griem~(\cite{ali_griem:errata}) theory which does not include the dispersive-inductive interactions.  

The new theory perhaps explains behaviour observed by Gardiner~et~al.~(\cite{gardiner}) and Castelli~et~al.~(\cite{castelli}) when using Balmer lines and Ali \& Griem~(\cite{ali_griem:errata}) theory to obtain effective temperatures for stars including the sun. It is superior to Ali \& Griem~(\cite{ali_griem:errata}) theory when applied to Balmer lines in the solar spectrum as it reduces the discrepancy between observed and computed profiles when the HOLMUL model is used and leads to the KNOVER model being classed as unacceptable both for its failure to adequately model the observed limb darkening and Balmer line profiles. However significant discrepancies between theory and observation for the Balmer lines still exist for the KOVER and HOLMUL models which could be due to our theory or the photospheric models.

Work is in progress on the determination of the effective temperatures of a sample of dwarf stars using our theory and observed Balmer lines for the stars in the sample. Preliminary results indicate a stronger correlation between the effective temperatures obtained from the Balmer lines and the effective temperature obtained by other methods such as the infrared flux method when our theory is used compared with that found when Ali \& Griem~(\cite{ali_griem:errata}) theory is used.

We plan to extend the theory to Paschen lines. We expect that in these lines dispersive-inductive interactions will dominate resonance interactions but not Stark broadening. Many of the approximations made in developing the theory are only valid in cool stars with weak quasistatic ion fields. For stars of earlier spectral type it may be necessary to develop a Unified Theory to correctly include self-broadening. The true limit of validity of the impact approximation in H$\alpha$ and H$\beta$ needs to be established.

\begin{acknowledgements}

We would like to thank Martin Asplund, Bengt Edvardsson, Bengt Gustafsson and Oleg Kochukhov for stimulating and encouraging discussions.  PB thanks Patrik Thor\'en for helping with MARCS model codes and Oleg Kochukhov for providing ATLAS models.  We would like to thank the referee, Professor T. Gehren, for constructive suggestions.  PB acknowledges the support of the Swedish Natural Science Research Council (NFR).  NSO/Kitt Peak FTS data used here were produced by NSF/NOAO.

\end{acknowledgements}

\end{document}